\documentclass[aps,pra,reprint,superscriptaddress,showpacs,amsmath,amssymb]{revtex4-1}

\usepackage[latin1]{inputenc}
\usepackage{graphicx}
\usepackage{amsmath}

\usepackage{amssymb}

\usepackage{stmaryrd}

\renewcommand{\r}{\mathbf{r}}
\renewcommand{\d}{\mathrm{d}}

\begin{document}


\title{Generalized Kirchhoff law } 

\author{Jean-Jacques Greffet$^{*}$ }
 \affiliation{Laboratoire Charles Fabry, Institut d'Optique Graduate School, CNRS, Universit\'e Paris-Saclay, 91127 Palaiseau,
France}
\author{Patrick Bouchon }
 \affiliation{ONERA-The French Aerospace Lab,
F-91761 Palaiseau, France} 
\author{Giovanni Brucoli }
 \affiliation{Laboratoire Charles Fabry, Institut d'Optique Graduate School, CNRS, Universit\'e Paris-Saclay, 91127 Palaiseau,
France}
\author{Emilie Sakat }
 \affiliation{Laboratoire Charles Fabry, Institut d'Optique Graduate School, CNRS, Universit\'e Paris-Saclay, 91127 Palaiseau,
France}
\author{Francois Marquier }
 \affiliation{Laboratoire Charles Fabry, Institut d'Optique Graduate School, CNRS, Universit\'e Paris-Saclay, 91127 Palaiseau,
France}


\begin{abstract} Thermal emission can be conveniently described using Kirchhoff law which states that the emissivity is equal to the absorptivity for isothermal bodies. For a finite size system, absorptivity is replaced by an absorption cross section. Here, we study the link between thermal emission and absorption by a finite size object which is not isothermal. We define a local absorption rate for a given incident plane wave and we prove that it is equal to the local emissivity rate. Hence, Kirchhoff law can be extended to anisothermal media. A practical consequence is the possibility of analysing thermal radiation by a variety of non-equilibrium systems such as microwave radiation in geophysical remote sensing or X-UV radiation by plasmas. This result provides a theoretical framework to analyse thermal emission by hot electrons in quantum wells, tunnel junctions or graphene. It paves the way to the design of a new generation of incandescent emitters made of subwavelength hot emitters coupled to cold antennas. The antennas control the emission spectrum, direction and polarization of the emitted radiation.

 \end{abstract}


\maketitle

\section{Introduction}

The purpose of this paper is to derive Kirchhoff law in the framework of fluctuational electrodynamics introduced by Rytov\cite{Rytov,Landau}. Kirchhoff law plays a fundamental role in the study of thermal emission. It allows to predict the thermal emission by an object from the knowledge of the absorption of an incident coherent plane wave. Yet, there is no general proof of its validity. The usual derivation given in textbooks is based on energy conservation. Hence, an additional detailed balance assumption is required to derive the equality of absorptivity and emissivity for a specified frequency, direction and polarization. There is a rigorous proof of its validity for an arbitrary sphere by Kattawar \cite{Kattawar}. It has also been proven within the scalar approximation for any planar interface separating vacuum from any complex medium satisfying reciprocity \cite{Greffet,Snyder}. Many works compare for different systems a direct calculation of the emission based on the fluctuational electrodynamics and a direct calculation of the absorption. So far, all these numerical calculations were found to agree with Kirchhoff law \cite{Joan1, Joan2, Zhang}. Yet, there is no general proof for an arbitrary finite size object. This unsatisfactory status of the derivation of Kirchhoff law has led to several works questioning its validity.  Of particular practical interest is the question of the existence of an upper bound to the emitted power. Is it possible to emit more than a black body ? Such an emitter has been called superplanckian emitter recently. For a finite size object, the power radiated can be modified by modifying its optical environment. It has been reported that the power emitted by a finite size object with area $A$ can be increased up to $n^2 A\sigma T^4$ by placing the emitter in a medium of refractive index $n$. The energy can then be transferred to the vacuum avoiding total internal reflection by using a solid immersion lens type of geometry as discussed in Refs \cite{Harrick,Fan}. This is in full agreement with standard radiometry as the thermodynamic radiance varies as $n^2I_b(T,\omega)$. There have been experimental reports and theoretical claims that thermal emission exceeding black body radiation in the far field is possible using a photonic crystal \cite{Lin}. This result was refuted \cite{Green} and subsequent work \cite{Fleming} reported an experimental bias. More recently, it has  been suggested that hyperbolic metamaterials could be used to achieve superplanckian emitters \cite{Nefedov,Maslovski}. This short literature survey shows that the existence of an upper bound of the thermal emission is still an open question. It is thus important to clarify the status of Kirchhoff law for finite size objects. 

Another important issue regarding thermal emission is the case of anisothermal bodies. In that case, Kirchhoff law cannot be applied. Obviously, the derivation based on thermodynamic equilibrium cannot be used to deal with anisothermal objects. However, it is possible to compute the emitted radiance through a direct calculation based on the fluctuational electrodynamics approach. This type of calculation has been reported in Refs \cite{Joan1,Joan2,Kong,Zhang, Han,Bardati,Yurasova} for instance. It can be used to solve the inverse problem in order to deduce the temperature field from the emitted radiance. Another application consists in controlling the emitted radiance. It has been proposed  \cite{Norris} to emit at different wavelengths or in different directions or polarizations by heating different parts of an object. Heating only submicron volumes of a structure could be an alternative to achieve modulation of the power emitted by an incandescent source faster than $100$ MHz \cite{Greffetnature,apl,Noda}. The question is therefore if it is possible to generalize Kirchhoff law to anisothermal systems. Recently, Han has derived a closed form of both the local absorption rate and the local emission rate and has proven their equality \cite{Han} in the case of a periodic multilayer system. Here, we generalize to any finite size body the approach taken by Han. We introduce a generalized Kirchhoff law establishing the equality between the local absorption rate and the local emission rate in any finite size body with arbitrary shape, orientation and structure. The only requirements are i) all materials satisfy reciprocity (i.e. they have a symmetrical permittivity tensor) and ii) it is possible to define a local temperature (local thermal equilibrium). This paves the way to the discussion of the engineering of the emission properties of a hot object surrounded by a cold structure operating as an antenna. Finally, we push the generalization of Kirchhoff law to anisothermal systems one step further by considering the case where different excitations (e.g. electrons, excitons, phonons) are at different temperatures at the same position. Thermal emission by hot electrons has been observed in a variety of systems such as tunneling tips \cite{Dumas, Bouhelier}, graphene \cite{Avouris,Hone} and quantum wells \cite{Sirtori}. The theoretical tools introduced in this paper provide a rigorous framework to analyze and optimize light emission by these systems.

\section{Generalized Kirchhoff law}
 To characterize the absorption of a linearly polarized monochromatic plane wave by a finite size body with volume $V$, it is useful to introduce the absorption cross-section $\sigma_{abs}$ which connects the absorbed power $P_{abs}$ with the incident Poynting vector flux:
 
 \begin{equation}
P_{abs}^{(l)}(\omega)=\sigma_{abs}^{(l)}(\mathbf{u}, \omega)\frac{\epsilon_0 c \vert  \mathbf{E}_{inc}^{(l)} (\omega)\vert^2}{2}.
 \label{eq1}
\end{equation}
The absorption cross-section depends on the polarization $l=s,p$, the incident direction $\mathbf{u}$ and the frequency $\omega$. We note that this concept does not provide any information about the position where absorption takes place in the body. An alternative form of the absorbed power is given by the integral over volume $V$ of the dissipation rate per unit volume:

\begin{equation}
P_{abs}^{(l)}(\omega)= \int _V \mathrm{Im}  [\epsilon(\mathbf{r'},\omega)]  \frac{\omega \epsilon _0}{2} \vert \mathbf{E}^{(l)} (\mathbf{r'},\omega)\vert ^2 d^3\mathbf{r'},
\label{eq:absorbed_power_general}
\end{equation}
where $\mathbf{E}^{(l)} (\mathbf{r'})$ is the field in the body illuminated by a $l$-polarized plane wave with incident field amplitude $E_{inc}$ and $\epsilon(\mathbf{r'},\omega)$ is the permittivity. This form describes explicitly where the absorption takes place but is not related explicitly to the incident field so that it cannot be expressed in terms of cross section. We now seek a connection between the incident field $\mathbf{E}_{inc}$ and the field in the absorber $\mathbf{E}(\mathbf{r})$. We will show that the existence of this linear relation allows to cast the absorption cross section in the form:

\begin{equation}
\sigma_{abs}^{(l)}(\mathbf{u},\omega)=\int_V \d^3\r' \alpha^{(l)} (\mathbf{u},\r',\omega),
\label{eq2}
\end{equation}
where $\alpha^{(l)}$ appears as an absorption cross section density. The unit vector $\mathbf{u}$ denotes the propagation direction of the incident plane wave (see Fig. 1). Note that just like the absorption cross section, this quantity depends on the object shape and orientation, it is not a material intrinsic property. Note in particular that the absorbing object may have electromagnetic modes which can be resonantly excited leading to enhanced absorption at some particular locations and frequencies. 

\begin{figure}[htb]
\centering
\includegraphics[width=80mm]{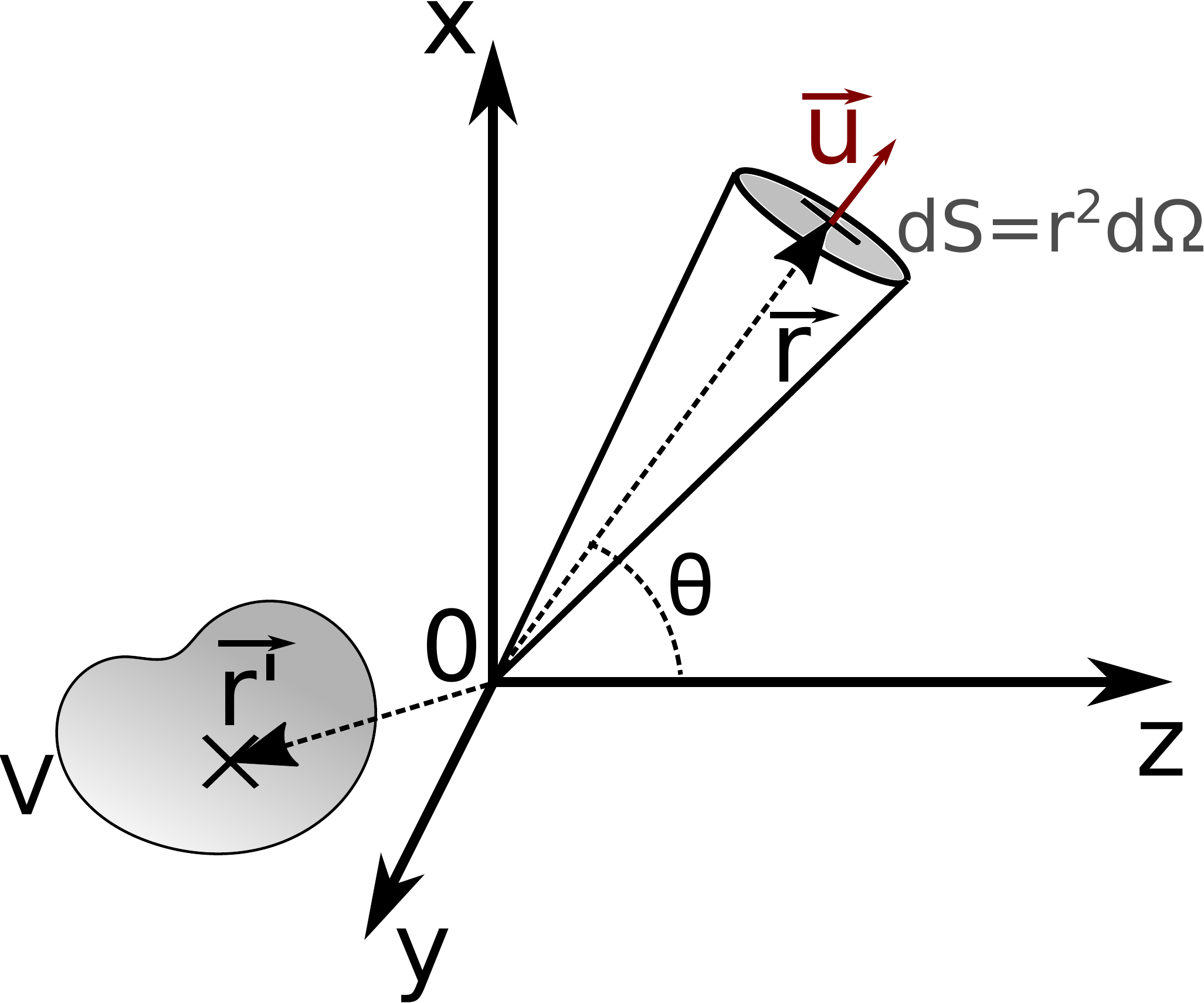}
\caption{ Sketch of the system. A finite size volume $V$ located in the half-space $z< 0$ radiates in the solid angle $\d \Omega$ subtended by the surface $\d S$.} 
\label{FIG1}
\end{figure}

We now turn to the power emitted by the object.  Using the fluctuational electrodynamics framework, we will write it as an integral over the volume of the emitter. We now define a local $(l)$ polarized emissivity density $\eta^{(l)}(\mathbf{u}, \r', \omega)$ using:

\begin{equation}
P_{e}^{(l)}=\int_{0}^{\infty} \d \omega \int _V \int_{4\pi} \eta^{(l)}(\mathbf{u}, \r', \omega) \frac{I_b [T(\r'),\omega] }{2} d^3\r' d \Omega,
\label{eq3}
\end{equation}
where we have introduced the blackbody radiance $I_b [T(\r'),\omega]=\frac{\omega ^2}{4 \pi ^3 c^2}\frac{\hbar \omega}{\exp(\hbar\omega/kT)-1}$. Note that we have introduced the polarized blackbody radiance $I_b/2$ which is half the blackbody radiance. 

 Both quantities, the absorption cross section density $\alpha^{(l)}$  and the emissivity density $\eta^{(l)}$ are polarized, directional and monochromatic. The first goal of this paper is to establish a general form of these two quantities and prove that they are equal for any body made of materials satisfying reciprocity, namely, materials with symmetric permittivities. This result is the generalized form of Kirchhoff law:
 
 \begin{equation}
 \eta^{(l)}( \mathbf{u}, \r', \omega)= \alpha^{(l)} (\mathbf{u},\r',\omega).
 \label{GKL}
 \end{equation}

An immediate consequence of this result is that the total power emitted by an isothermal object in the solid angle $\d \Omega$ is given by:
\begin{equation}
\d P_e^{(l)}= \int_{0}^{\infty} \d\omega \; \sigma^{(l)}_{abs}(\mathbf{u},\omega) \;\frac{ I_b(T,\omega)}{2} \; \d\Omega.
\end{equation}

This result provides all the required information about the emission and shows that it is entirely characterized by the knowledge of the absorption cross-section. We emphasize that it is valid for any size of the object. We also emphasize that the absorption cross section is not related to the actual geometrical size of the object. We can thus consider that the particle has an effective area called the absorption cross section $\sigma^{(l)}_{abs}(\mathbf{u},\omega)$  that can be used to characterize both emission and absorption. In that sense, the two are always equal and there is no superplanckian emission in the far field: the body can be considered to have an effective emissivity or absorptivity which is equal to 1. The concept of superplanckian emission can be introduced if one compares the absorption cross section to the actual geometrical section of the object in a situation where the geometrical section is smaller than the absorption cross section. It is well known that the absorption cross section of a resonant subwavelength sphere can be much larger than the geometrical cross section \cite{Bohren}. It is also known that in the so-called resonant regime, namely for sizes on the order of the wavelength, the absorption cross section can be larger than the geometrical cross section. That would correspond to the so-called super-planckian emitters. However, in the subwavelength and resonant regimes, geometrical optics is not valid so that i) absorption is not expected to be proportional to the geometrical area, ii) the concepts of radiometry such as emissivity and absorptivity are not expected to be valid. Hence, in what follows, we only use the absorption cross section concept which is unequivocally defined as the response of an object to a plane wave. Finally, as opposed to the Kirchhoff law which is only valid for isothermal bodies, the generalized Kirchhoff law can be used to compute the power emitted by an anisothermal body. Here, we will derive the emitted power in terms of the absorption cross section density:

\begin{equation}
\d P_e^{(l)}= \int_{0}^{\infty}\d\omega \int_V \d^3\r'  \alpha_{abs}^{(l)} (\mathbf{u},\r',\omega) \frac{ I_b[T(\r'),\omega]}{2}\d\Omega.
\label{Pe}
\end{equation}

 In the rest of the paper, we proceed to derive the generalized Kirchhoff law using the reciprocity of a Green tensor. In the first part of the paper, we derive the emission cross section density. In the second part of the paper, we derive the absorption cross section density. Both are given in terms of the Green tensor and the imaginary part of the dielectric permittivity. From a practical point of view, the explicit forms of $\eta^{(l)}$ and $\alpha^{(l)}$ are rather cumbersome. In practice, it is possible to compute numerically $\alpha^{(l)} (\mathbf{u},\r',\omega)$ and insert it in Eq.(\ref{Pe}) to derive the emitted power.

\begin{figure}[htb]
\centering
\includegraphics[width=80mm]{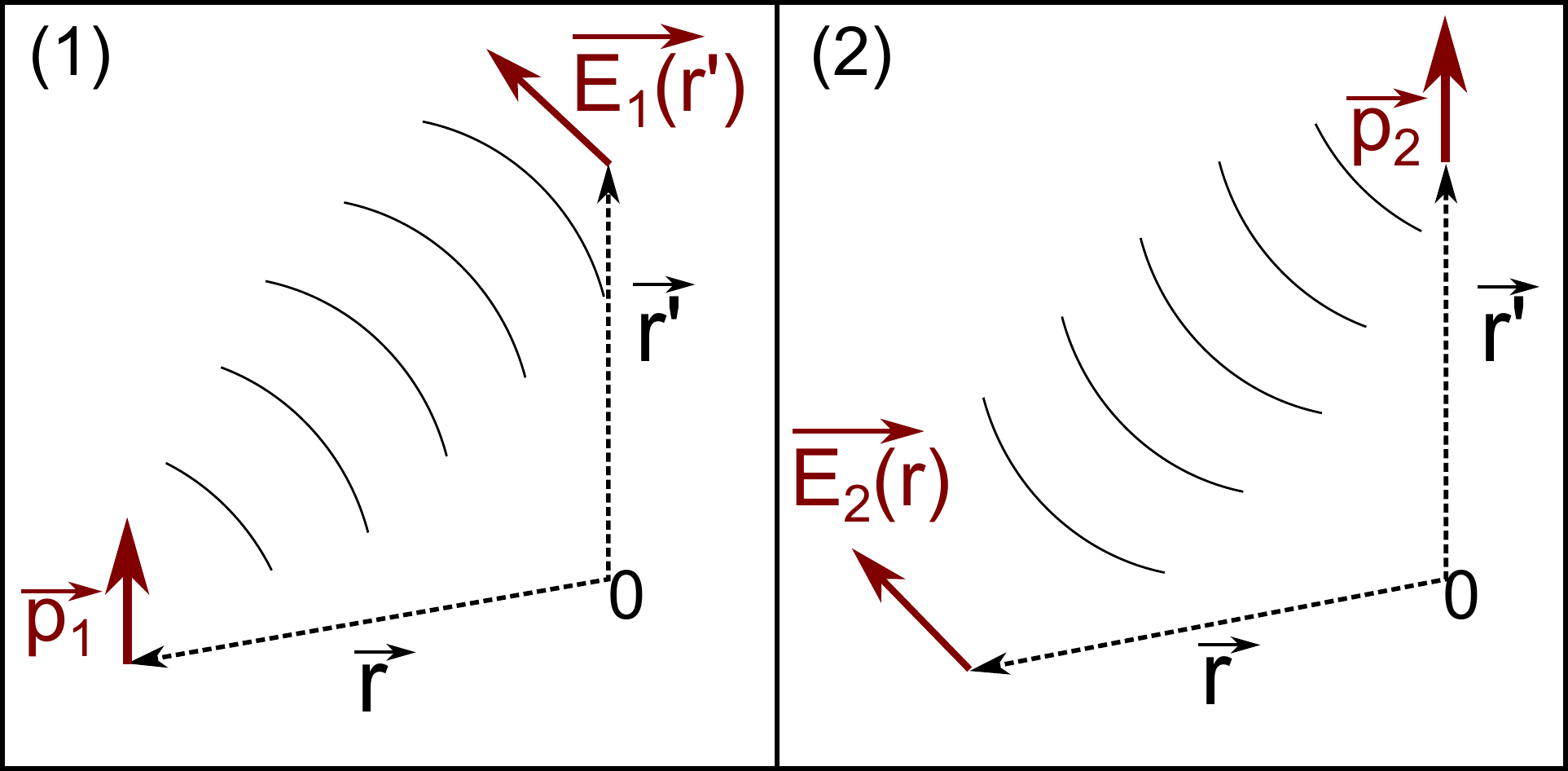}
\caption{ Illustration of two reciprocal situations (1) and (2) obtained upon exchanging source and detector positions.} 
\label{FIG2}
\end{figure}

Before proceeding to the formal derivation, we will emphasize two basic ingredients. The first key point is the reciprocity of the Green tensor. This is better understood by looking at Figure 2. The basic idea is that a dipole moment $\mathbf{p}_1$ located at $\mathbf{r}$ generates a field $\mathbf{E}_1$ at $\mathbf{r'}$. Conversely, a dipole moment  $\mathbf{p}_2$ located in the absorber at $\mathbf{r}'$ generates a field $\mathbf{E}_2$ at the position $\mathbf{r}$. It follows from reciprocity that these two fields satisfy $\mathbf{p}_1\cdot\mathbf{E}_2=\mathbf{p}_2\cdot\mathbf{E}_1$ \cite{Greffet97}. In others words, the signal is not changed by exchanging point-like sources and detectors. The second key ingredient in the generalized Kirchhoff law is the introduction of the emission angle $\theta$. To proceed, we will use the asymptotic form of the field in the far field. We now turn to the derivation of the local emission and absorption rates.

\section{Emission}

We begin by calculating the thermal emission from a local point $\mathbf{r'}$ towards a point $\mathbf{r}$. To proceed, we consider that thermal fields are radiated by stationary random currents as discussed in \cite{Rytov,Landau}. In what follows we use the notation of Ref. \cite{Landau} for the spectral analysis of stationary random processes. We use the correlation function of the current density given by the fluctuation-dissipation theorem for a medium which is linear and isotropic:
\begin{eqnarray}
&\langle j_n(\r,\omega)j_m(\r',\omega')\rangle = \nonumber \\
&2\pi\delta(\omega+\omega')\delta(\r-\r')\delta_{nm}2\omega\epsilon_0\mathrm{Im}[\epsilon(\r',\omega)]\Theta[T(\r'),\omega],
\label{FD}
\end{eqnarray}
where the brackets denote ensemble average, $\Theta[T(\r'),\omega]=\hbar\omega/(\exp(\hbar\omega/k_BT(\mathbf{r}'))-1)$ and $T(\mathbf{r}')$ is the temperature at point $\mathbf{r}'$.
We consider a point $\mathbf{r}$ in the far field so that the emission direction is specified by the unit vector $\mathbf{u}=\mathbf{r}/r$.  The electric field is transverse to the propagation direction $\mathbf{u}$ so that it can be described using only two components. Let us introduce two orthogonal unit vectors denoted $\mathbf{e}^{(s)}$ and $\mathbf{e}^{(p)}$ perpendicular to $\mathbf{u}$. The emitted power flowing through the area $\d A$ is given by the flux of the Poynting vector:
\begin{equation}
\d P_e=\langle\mathbf{E}(\mathbf{r},t)\times\mathbf{H}(\mathbf{r},t)\rangle\cdot \d A\mathbf{u}.
\end{equation}
Let us formally introduce the Fourier transform of the electric field:
\begin{equation}
\mathbf{E}(\mathbf{r},t)=\int_{-\infty}^{\infty}\frac{\d \omega}{2\pi}\mathbf{E}(\mathbf{r},\omega)\exp(-i\omega t).
\end{equation}
In the far field, the electromagnetic field has a plane wave structure so that $\mathbf{H}(\mathbf{r},\omega)=\epsilon_0 c \mathbf{u}\times \mathbf{E}(\mathbf{r},\omega)$.
It follows that the Poynting vector through $\d A$ can be cast in the form:
\begin{eqnarray}
\d P_e=\d A\; \epsilon_0 c \int_{-\infty}^{\infty}\frac{\d \omega}{2\pi}\int_{-\infty}^{\infty}\frac{\d \omega'}{2\pi}\exp[-i(\omega+\omega')t]  \nonumber \\
\langle\mathbf{E}(\mathbf{r},\omega)\cdot\mathbf{E}(\mathbf{r},\omega')\rangle .
\label{emission}
\end{eqnarray}

The $m-$component of the electric field is given by:
\begin{equation}
E_m (\r,\omega)=  i \mu _0 \omega\int G_{mn}(\mathbf{r},\mathbf{r'},\omega)j_n(\mathbf{r'},\omega)\d^3\mathbf{r'},
\label{eq:eq_champE_green}
\end{equation}
where $G_{mn}$ is a component of the Green tensor and $j_n$ is the $n-$component of the current density. Throughout the paper, we use the Einstein notation so that there is a sum over repeated indices. The amplitude of the field along the unit vector $\mathbf{e}^{(l)}$ is given by:
\begin{equation}
E^{(l)}=\mathbf{e}^{(l)}\cdot\mathbf{E}=i \mu _0 \omega\int e_m^{(l)}G_{mn}(\mathbf{r},\mathbf{r'},\omega)j_n(\mathbf{r'},\omega)d^3\mathbf{r'}.
\label{Green}
\end{equation}

After inserting Eq.(\ref{FD}) and Eq.(\ref{Green}) into Eq.(\ref{emission}) and using $G_{mn}(\r,\r',-\omega)=G_{mn}^*(\r,\r',\omega)$,  we get the power emitted in $l$-polarization through $\d A$:

\begin{eqnarray}
\d P_e^{(l)}=&\d A \int_{-\infty}^{\infty}\frac{\d \omega}{2\pi}   e_m^{(l)}G_{mk}(\mathbf{r},\mathbf{r'},\omega)e_n^{(l)}G_{nk}(\mathbf{r},\mathbf{r'},-\omega)  \nonumber \\
& 2 k^3 \mathrm{Im}[\epsilon(\r')]\Theta[T(\r'),\omega].
\label{Pe_0}
\end{eqnarray}

We now evaluate the power emitted per unit solid angle.
We start by introducing a plane wave expansion \cite{Mandel} of the Green tensor at the observation point $\r$:

\begin{equation}
G _{mn}(\r,\r',\omega)= \int G _{mn}(\mathbf{k}_{\parallel},z=0,\r',\omega) 
e^{i\mathbf{k}_{\parallel}\cdot\r}e^{i \gamma z}\frac{d^{2}\mathbf{k}_\parallel}{(2\pi)^2},
\label{eq:eq_asymp_0}
\end{equation}
where $\gamma=[\omega^2/c^2-k_{\parallel}^2]^{1/2}$,  $\mathbf{k}_{\parallel}=(k_x,k_y,0)$ is the wavevector in the (x,y) plane and $G _{mn}(\mathbf{k}_{\parallel},z=0;\r';\omega) $ is the Fourier transform of $G _{mn}(\r,\r',\omega)$ with respect to $x$ and $y$ in the plane $z=0$. An asymptotic evaluation of this integral when $kr \rightarrow \infty$ by the method of the stationary 
phase \cite{Mandel} gives:
\begin{equation}
G _{mn}  (\r,\r',\omega)\rightarrow \frac{-i k}{2\pi} \cos \theta \,
G _{mn}(\mathbf{k}_{\parallel},z=0,\r',\omega)\frac{e^{ikr}}{r}
\label{eq:eq_asymp}
\end{equation}

Inserting this asymptotic form in Eq.(\ref{Pe_0}) and introducing the solid angle $\d \Omega=\d A/r^2$ yields:

\begin{eqnarray}
\d P_{e}^{(l)}=&\int_{-\infty}^{\infty}\frac{\d \omega}{2\pi} \frac{k^5}{2\pi^2}\cos ^2 \theta \,
 \int \vert  e_m^{(l)} G_{mn} (\mathbf{k}_{\parallel},z=0,\r',\omega)\vert ^2 \nonumber \\
&\mathrm{Im}[\epsilon(\r'),\omega] \Theta (T(\r',\omega))  \d^3\r' \d \Omega.
\label{eq:emitted_power}
\end{eqnarray}

We now cast the result in a form that can be compared to the standard radiometric approach. We restrict the integration over positive frequencies. The emitted power can be cast in the form:

\begin{equation}
dP_{e}^{(l)}=\int_{0}^{\infty} \d \omega \int _V  \eta^{(l)}( \mathbf{u}, \r',\omega) \frac{I_b [T(\r'),\omega] }{2} \d^3\r' \d \Omega,
\label{eq:emitted_power2}
\end{equation}

where we have defined the polarized spectral directional emissivity $\eta^{(l)} (\mathbf{u},\r',\omega)$:
\begin{eqnarray}
&\eta^{(l)} (\mathbf{u}, \r',\omega) = \nonumber \\
& 4 k^3 \cos^2 \theta 
 \vert e_m ^{(l)} G_{mn}(\mathbf{k}_{\parallel}, z=0,\r',\omega)\vert ^2 \mathrm{Im}(\epsilon(\r'),\omega).
\label{eq:emissivity}
\end{eqnarray}

Note that we define the polarized emission as proportional to $\frac{I_b [T(\r'),\omega] }{2}$ so that the total emitted power is proportional to $(\eta^s+\eta^p)\frac{I_b [T(\r'),\omega] }{2}$.

\section{Absorption}

To proceed, we consider that the incident electric field is generated by a point-like dipole located ar $\mathbf{r}$ in the far field with a $l$-polarized dipole moment $p_{inc}\mathbf{e}^{(l)}$. The $m-$component of the field $\mathbf{E}^{(l)}(\r')$ generated at $\mathbf{r}'$ is then given by:

\begin{equation}
 E ^{(l)}_m (\r')=  G_{mn}(\r',\r,\omega) e^{(l)}_n p_{inc} \mu _0 \omega ^2,
\label{eq:green_tensor_def}
\end{equation}

and the amplitude of the incident field produced by the $l-$polarized electric dipole at point $\mathbf{r}$  is given by:

\begin{equation}
E_{inc}^{(l)}=\frac{\exp(ikr)}{4\pi r}\mu_0 \omega^2 p_{inc}.
\label{eq:E-p}
\end{equation}

It follows that:
\begin{equation}
 E ^{(l)}_m (\r')=  G_{mn}(\r',\r,\omega) e^{(l)}_n 4\pi r E_{inc}^{(l)} \exp(-ikr)
\label{eq:E_l}
\end{equation}

Besides, we can replace the Green tensor using the reciprocity theorem:

\begin{equation}
G_{mn}(\r',\r,\omega)=G_{nm}(\r,\r',\omega),
\label{eq:thm_recipro}
\end{equation}

to obtain:

\begin{equation}
P_{abs}^{(l)}= \int _V \mathrm{Im}[\epsilon(\mathbf{r'},\omega)]  \frac{\omega \epsilon _0}{2} \vert   e^{(l)}_n G_{nm}(\r,\r',\omega)  4\pi r E_{inc}  \vert ^2 d^3\mathbf{r'}.
\label{eq:absorbed_power_general2}
\end{equation}

We finally insert the already used 
asymptotic evaluation of the Green tensor (\ref{eq:eq_asymp}) to get :

\begin{equation}
P_{abs}^{(l)}=  \frac{\epsilon _0 c}{2}  \vert E_{inc} \vert ^2
  \int _V d^3\mathbf{r'} \alpha^{(l)} (\mathbf{u},\r',\omega)
\label{eq:absorbed_power}
\end{equation}

where we have defined a polarized directional absorption cross section density $\alpha^{(l)} (\mathbf{u},\r',\omega)$:
\begin{equation}
\alpha^{(l)} (\mathbf{u},\r',\omega) =   4 k^3  \mathrm{Im}[\epsilon(\mathbf{r'})]  \cos^2 \theta 
\vert   e^{(l)}_n G_{nm}(\mathbf{k}_{\parallel},z=0,\r',\omega)  \vert ^2
\label{eq:absorptivity1}
\end{equation}

Upon inspection, we see that $$\eta^{(l)} (\mathbf{u},\r',\omega)=\alpha^{(l)} (\mathbf{u},\r',\omega),$$
which is the generalized form of the Kirchhoff law.

\section{Anisotropic media}

In this section, we extend the derivation to the case of anisotropic media. It is known that some anisotropic materials may display unusual optical properties associated to hyperbolic dispersion relations. It has been suggested that these features may have implications for heat transfer \cite{Nefedov, Maslovski}. Here, we stress that there are no consequences on the validity of the generalized Kirchhoff law provided that the system is composed of materials with symmetric permittivity tensors $\epsilon_{nm}=\epsilon_{mn}$ as required by reciprocity. The form of the local absorption/emission rate is different but absorption rate and emission rate are still equal. Using the relevant form of the fluctuation-dissipation theorem:
\begin{eqnarray}
&\langle j_n(\r,\omega)j_m(\r',\omega')\rangle =2\pi\delta(\omega+\omega')\delta(\r-\r') \nonumber \\
&2\omega\epsilon_0\mathrm{Im}[\epsilon_{nm}(\r',\omega)]\Theta[T(\r'),\omega],
\label{eq:FD2}
\end{eqnarray}
we find
\begin{eqnarray}
&\eta^{(l)} (\mathbf{u},\r',\omega)=\alpha^{(l)} (\mathbf{u},\r',\omega) =   4 k^3 \cos^2 \theta \,\mathrm{Im}[\epsilon_{pq}(\mathbf{r'},\omega)]  \nonumber \\
&e^{(l)}_n G_{np}(\mathbf{k}_{\parallel},0,\r',\omega)  e^{(l)}_m G_{mq}(\mathbf{k}_{\parallel},0,\r',\omega).
\label{eq:absorptivity2}
\end{eqnarray}

\section{Emission by anisothermal systems}

The generalized Kirchhoff law introduced above allows a direct calculation of the power emitted by anisothermal systems. An example is the earth infrared radiation due to anisothermal soils \cite{Kong,Bardati}. Another example at a very different length scale is a graphene film deposited on a substrate \cite{Avouris,Hone}. When the current density in graphene is very large, the electronic temperature can be increased above 1000 K while the substrate remains at lower temperatures. A similar example is the radiation produced by hot electrons in a in a metallic tip or in a quantum well \cite{Dumas, Bouhelier,Sirtori}. These examples are interesting as they introduce another feature: it is possible to define a temperature for the electrons and a temperature for the lattice. This is the so-called two temperatures model. In that case, we have two different temperatures at the same point but for different systems. It is possible to include this feature in the model by using the following form of the power spectral density of the current:

\begin{eqnarray}
&\langle j_n(\r,\omega)j_m(\r',\omega')\rangle = 2\pi\delta(\omega+\omega')\delta(\r-\r')2\omega\epsilon_0 \nonumber \\
& \{\mathrm{Im}[\epsilon_{el}(\r',\omega)]\Theta[T_{el}(\r'),\omega]+\mathrm{Im}[\epsilon_{la}(\r',\omega)]\Theta[T_{la}(\r'),\omega]\},\nonumber \\
\label{FD_2}
\end{eqnarray}

where $\epsilon_{el}$ ($\epsilon_{la}$) denote the electron (lattice) contribution to the permittivity and  $T_{el}$ ($T_{la}$) denote the electron (lattice) temperature. 

In summary, it is seen that if one can assign to each type of excitation (electrons, excitons, phonons) both a temperature and a contribution to the imaginary part of the permittivity, it is  possible to use the generalized Kirchhoff law to account for thermal radiation by non-equilibrium systems.\\

\section{Concluding remarks}

In summary, we have derived a generalized Kirchhoff law valid for any finite size object. When we apply it to isothermal objects, we find that the emission is always equal to the product of the blackbody radiance and the absorption cross section. Optimizing the emission is thus equivalent to optimizing the absorption cross-section. We note that the usual Kirchhoff law, i.e. the equality between the absorptivity and emissivity of an interface, appears as a particular case of the result derived here when we consider planar and isothermal objects much larger than the wavelength.  The generalized Kirchhoff law can be used for anisothermal objects including the case where different excitations are at different temperatures. 

Let us now explore the implications of the generalized Kirchhoff law for harnessing thermal radiation. It has been known for applications such as bolometers that absorption by a small volume of absorbing material can be increased using antennas \cite{Boreman}. Indeed, the antenna can capture more efficiently the incident power and funnel it into the absorber volume. This  absorbed power in the presence of the antenna is then proportional to an effective absorption cross section denoted $\sigma_{ant}$. In addition, the antenna can be directional and frequency selective \cite{Boreman}. It follows from the generalized Kirchhoff law that if the absorption in the absorber volume is enhanced then its thermal emission is enhanced. The total emitted power can be increased by the same factor $\sigma_{ant}/\sigma_{abs}$ which can be larger than two orders of magnitude. Furthermore, the emission can be directional and frequency selective. We anticipate that a metasurface consisting of a periodic  array of subwavelength hot objects connected to antennas could be optimized to behave as a blackbody antenna with unity emissivity while using only a very reduced amount of hot material. This approach paves the way to a new class of THz and IR thermal emitters with a controlled emission direction, spectrum and polarization.

\acknowledgements
We acknowledge the support of the Agence Nationale de la Recherche through the grant ANR-14-CE26-0023-03 and the support of the ONERA through the project SONS. This work was supported by the US Department of Energy, Office of Basic Energy Sciences, Division of Materials Sciences and Engineering. It was performed, in part, at the Center for Integrated Nanotechnologies, an Office of Science User Facility operated for the US Department of Energy (DOE) Office of Science. JJG is a senior member of the Institut Universitaire de France.\\

\clearpage

\end{document}